\documentclass[journal]{IEEEtran}
\ifCLASSINFOpdf
\else
\fi
\usepackage{graphicx}  
\usepackage{bm}        
\usepackage{amssymb}   
\usepackage{amsmath}
\usepackage{float}
\usepackage{color}
\usepackage{physics}
\usepackage{xparse,xcolor}

\begin{document}
%
\title{Branching optical frequency transfer with enhanced post automatic phase noise cancellation}
%
%
%

\author{Ruimin Xue, Liang Hu,~\IEEEmembership{Member,~IEEE,} Jianguo Shen, Jianping Chen and Guiling Wu,~\IEEEmembership{Member,~IEEE}
\thanks{Manuscript received xxx xxx, xxx; revised xxx xxx, xxx. This work was supported by the National Natural Science Foundation of China (NSFC) (61627871, 61905143). (Corresponding author: Liang Hu)}
\thanks{R. Xue, L. Hu, J. Chen and G. Wu are with the State Key Laboratory of Advanced Optical Communication Systems and Networks, Shanghai Institute for Advanced Communication and Data Science, Shanghai Key Laboratory of Navigation and Location-Based ServicesDepartment of Electronic Engineering, Shanghai Jiao Tong University, Shanghai 200240, China (e-mail: rmxue96@sjtu.edu.cn; liang.hu@sjtu.edu.cn;  jpchen62@sjtu.edu.cn; wuguiling@sjtu.edu.cn).}
\thanks{J. Shen is with the College of Physics and electronic information Engineering, Zhejiang Normal University, Jinhua, 321004, China (e-mail: shenjianguo@zjnu.cn).}
}

%
%

\markboth{JOURNAL OF LIGHTWAVE TECHNOLOGY,~Vol.~xxx, No.~xxx, xxx~2021}%
{Shell \MakeLowercase{\textit{et al.}}: Bare Demo of IEEEtran.cls for IEEE Journals}
%



\maketitle

\begin{abstract}
We present a technique for coherence transfer of laser light through a branching fiber link, where the optical phase noise induced by environmental perturbations via the fiber link is passively compensated by remote users without the requirements of any active servo components. At each remote site, an acousto-optic modulator (AOM) is simultaneously taken as a frequency distinguisher for distinguishing its unique frequency from other sites' and as an optical actuator for compensating the phase noise coming from the optical fiber. With this configuration,  we  incorporate a long outside loop path consisting of a fiber-pigtailed AOM into the loop, enabling the significant reduction of the outside loop phase noise in the passive way. To further address the residual out-of-loop phase noise coming from the interferometer and the two-way optical frequency comparison setup, we design a low-noise active temperature stabilization system. Measurements with a back-to-back system show that the stability  in our stabilization system is $2\times10^{-16}$ at 1 s, reaching $2\times10^{-20}$ after 10,000 s. Adopting these techniques, we demonstrate transfer of a laser light through a branching fiber network with 50 km and 145 km two fiber links. After being compensated for the 145 km fiber link, the relative frequency instability is $3.4\times10^{-15}$ at the 1 s averaging time and scales down to $3.7\times10^{-19}$ at the 10,000 s averaging time. This proposed technique is suitable for the simultaneous transfer of an optical signal to a number of independent users within a local area.


\end{abstract}


%
\IEEEpeerreviewmaketitle

\section{Introduction}




\IEEEPARstart{O}{ptical}-atomic clocks, which are among the most precise metrological instruments currently available, provide continuous-wave (CW) laser radiation that is stabilized to a narrow-linewidth atomic transition \cite{bloom2014optical, huntemann2016single, mcgrew2018atomic, oelker2019demonstration, brewer2019al+}. Optical frequency transfer over optical fibers provides an ideal solution to coherently transfer the stability of an optical clock to remote sites \cite{ma1994delivering, Predehl441, Calonico2014}. This optical-frequency transfer capability was developed for optical clock comparison and timekeeping \cite{takano2016geopotential, lisdat2016clock, guena2017first, grotti2018geodesy, collaboration2021frequency}. One of the most common ways to implement optical frequency transfer is the active phase noise cancellation method, in which the fiber phase noise is detected by comparing the local optical signal with the round-trip signal and the compensator is implemented by actuating the frequency of the transferring optical signal \cite{ma1994delivering}. In many applications, multiple-node optical frequency transfer is usually required \cite{grainge2017square, schediwy2019mid, he2018long, clivati2020common}. To address this dilemma, a pionnering work was demonstrated that can provide coherent optical-frequency signals at intermediate sites along a bus topology optical fiber link \cite{grosche2014eavesdropping, gao2012fiber, hu2020allpassive}. However, this approach is limited to the fiber link with a bus topology. By contrast, many applications require the transfer of signals from a single master site to multiple remote sites over a fiber network with a branching topology such as the planned Square Kilometer Array \cite{grainge2017square, schediwy2019mid, he2018long, clivati2020common}. An excellent scheme and its extensions \cite{schediwy2013high, wu2016coherence, ma2018delay} for fiber phase noise cancellation at the remote sites have been proposed and experimentally demonstrated.  In these schemes, the phase noise for each branching fiber link is obtained at the remote site by heterodyne beating the first-pass fiber output light against the light that transfers through the fiber for three times. The phase noise of the single-pass fiber output light can be compensated by using the detected phase noise with the active and passive phase noise cancellation techniques \cite{schediwy2013high, wu2016coherence, hu2020cancelling, hu2020allpassive}. In these schemes,  transfer to each remote site is simultaneous and independent, addressing the performance entanglement of all the remote sites of the after-mentioned multiple-access optical frequency transfer techniques \cite{grosche2014eavesdropping, gao2012fiber}. This has the advantage that if a single stabilization servo fails then only a single remote site will lose its stable signal. In our previous work, we have demonstrated optical frequency transfer with passive phase stabilization over a branching fiber network \cite{hu2020multi}. The technique possesses the advantages of an unlimited compensation precision and a fast compensation speed and free from the effect of servo bumps on the spectral purity \cite{hu2020cancelling, hu2020multi, hu2020allpassive} as opposed to the active phase noise compensation based optical frequency transfer technique \cite{schediwy2013high, wu2016coherence}.


In any optical frequency transfer systems, it is crucial to isolate the uncommon path well from the laboratory environment as any perturbations on the uncommon path will be recognized as the phase noise.  Optical frequency transfer with the uncertainty as low as a few $10^{-21}$ was reported with various experimental implementation \cite{bercy2014two, raupach2015brillouin, bercy2016ultrastable}.  Among them, the biggest one arises from the so-called interferometric noise \cite{williams2008high, stefani2015tackling}. The intrinsic interferometric noise was experimentally observed and reported \cite{bercy2016ultrastable, stefani2015tackling, xu2018studying}. Williams \textit{et al.} \cite{williams2008high} reported this instability floor as $2\times10^{-17}$ at 1 s, scaling down with the slope of $1/\sqrt{\tau}$, and reaching $2\times10^{-19}$ at 10,000 s. The noise floor of the interferometer is strongly related to temperature drifts acting on the uncommon arm that is the out-of-loop path \cite{bercy2016ultrastable, stefani2015tackling, pinkert2015effect, tian2020hybrid}.  The relative phase variations $y$ with the uncommon path length of $\delta l$ due to the time-varying temperature $T(t)$ can be expressed as \cite{bercy2016ultrastable, stefani2015tackling, pinkert2015effect, tian2020hybrid},
\begin{equation}
y =\frac{\Delta\nu}{\nu}=\frac{1}{c}\left(\delta l\frac{dn}{dT}+n\frac{dl}{dT}\right)\frac{dT(t)}{dt}
\end{equation}
where $c$ is the speed of light, $\nu$ is the optical carrier frequency and $\Delta\nu$ is the absolute frequency shift. In our previous branching optical frequency transfer with the remote phase noise cancellation, the phase noise detected by the remote site is passively compensated by cascading one more acousto-optic modulator (AOM) just after the remote AOM \cite{hu2020multi}. With this configuration, it significantly increases the outside loop uncommon path.  For a typical temperature perturbation with the sinusoidal fluctuation amplitude and cycle of 1 K and 3,600 s, respectively, one expects  overlapping Allan deviation (OADEV) as high as $4\times10^{-18}$ at 1,800 s for the out-of-loop up to $\delta l=0.1$ m, which is consistent with the previous experimental results \cite{hu2020multi}.

To suppress this effect, Stefani \textit{et al.} demonstrated the interferometer temperature sensitivity can be effectively reduced from 7 fs/K to 1 fs/K by actively stabilizing the temperature of the interferometer \cite{stefani2015tackling}. The phase noise can be further reduced when the optical fiber was installed in an ultrastable environment with evacuation, vibration isolation, acoustic shielding, and temperature stabilization as demonstrated in \cite{wada2018evaluation}. The temperature effect can be further suppressed by acting a temperature measurement on the uncommon arms of the interferometers \cite{xu2019reciprocity}, allowing to compensate for the interferometric noise using the post-processing techniques. As the refractive of the air is the three orders of magnitude lower than that of the fiber, alternative way to suppress the interferometer noise is to build the interferometer with free-space optics \cite{kang2019free}. The free-space interferometers have been widely demonstrated \cite{williams2008high, ma2015coherence, hu2020passive}. However, they are sensitive to the air flow and the long-term stability is still limited to the $10^{-20}$ level \cite{williams2008high, ma2015coherence, hu2020passive}, so that the precise temperature stabilization system is still necessary to perform the better long-term stability \cite{williams2008high, ma2015coherence, hu2020passive}.  Furthermore, the high cost and low reliability of the free-space interferometers will become a great challenge for the field-deployed application.  Recently, Akatsuka \textit{et al.} reported laser repeater stations with photonic planner lightwave circuits (PLC), in which all the passive components including the couplers, waveplates and polarization beam splitters are successfully integrated into a single PLC circuit, illustrating that the interferometer noise floor was evaluated to be $1\times10^{-18}$ at 1 s and $1\times10^{-21}$ at 5,000 s \cite{akatsuka2020optical}. In our group, we also demonstrated an optical frequency transfer system by integrating the interferometer into the silicon chips \cite{liang2021towards}.

All the above techniques used for improving the noise floor of the interferometer have been done by shortening the out-of-loop path length or stabilizing the temperature of the path. We believe that the most promising way to  improve the noise floor of the interferometer is to incorporate the out-of-loop components into the loop in such a way that the in-loop signal experiences exactly twice the perturbations. In this way, the out-of-loop components will become part of the stabilized fiber link and the phase noise coming from these components can be effectively suppressed. By doing this way, it also reduces the requirement of the temperature stabilization setup. Beside the noise from the interferometer itself, the ability to measure the end-to-end stability of the optical frequency transfer system is another important issue that needs to be addressed. A common way to achieve this is to connect the outputs of the two sites with an equivalent arm length $1\times2$ optical coupler and the beatnote of the combined lights is recovered by a photon-detector (PD). However, this configuration still introduces the uncommon optical path.  A simple way to avoid the outside path for evaluating the system is to put the local site and remote site at the same housing as adopted in \cite{grosche2009optical, williams2008high}. However, this configuration can not be directly applied in the field application, in which the two sites are separated far away.

\begin{figure*}[htbp]
\centering
\includegraphics[width=1\linewidth]{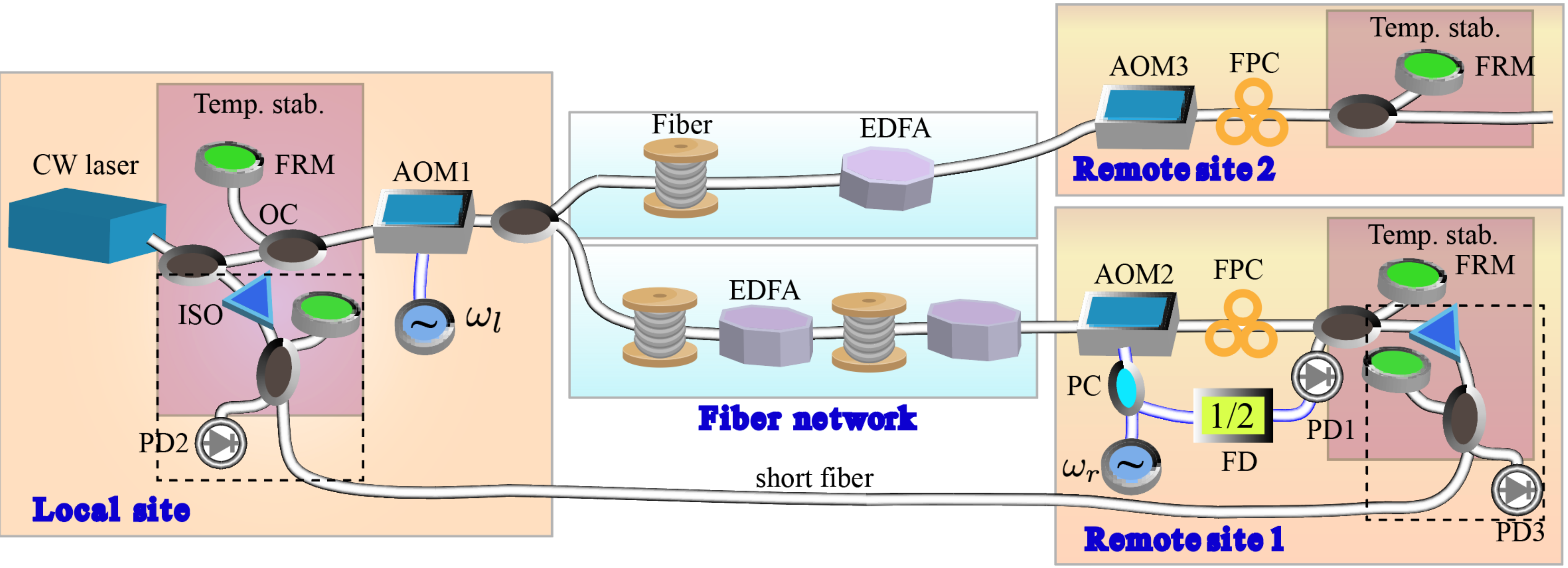}
\caption{Schematic diagram of our enhanced passive optical frequency transfer on a branching fiber network, where the optical phase noise induced by environmental perturbations via the fiber link is passively compensated by remote users. At each remote site, only single  acousto-optic modulator (AOM) is simultaneously used as a frequency distinguisher for distinguishing its unique frequency from other sites' and as a compensator  for compensating the phase noise coming from the optical fiber. The system performance is characterized by the two-way optical frequency comparison method as shown by the dashed boxes. The uncommon fiber components at each site are housed in a two-layer temperature stabilized box with the peak-to-peak fluctuations of 0.01 K. Here the schematic diagram of the remote site 2 is simplified because it has the same configuration with the remote site 1. OC: optical coupler, ISO: isolator, AOM: acousto-optic modulator, FD: radio-frequency divider, PC: power combiner, PD: photo-detector, EDFA: erbium-doped fiber amplifier, FRM: Faraday mirror, FPC: fiber polarization controller.}
\label{fig1}
\end{figure*}

In this article, we propose a novel technique for coherence transfer of laser light through a branching fiber link, where the optical phase noise induced by environmental perturbation via the fiber link is passively compensated by remote users as opposed to the active one adopted in \cite{schediwy2013high}. To incorporate the outside loop components into the loop  such as the fiber-pigtailed acousto-optic modulator (AOM) \cite{hu2020multi}, allowing to reduce the length of the out-of-loop path, at each remote site only single  AOM is simultaneously used as a frequency distinguisher for distinguishing its unique frequency from other sites' and as a phase compensator  for compensating the phase noise coming from the optical fiber. Furthermore, in comparison with multiple-user optical frequency transfer over a ring and a bus fiber links \cite{hu2020allpassive, hu2020passive},  in which the stabilization of each user end is strongly dependent on the main fiber link, high-performance multiple-user optical frequency transfer demonstrated here not only increases the adaptability of optical frequency transfer to the existing fiber network, but also improves the reliability of the network because the phase noise compensation of each user node is independent of each other. Moreover, the two-way optical frequency comparison method was used to measure the end-to-end stability for avoiding uncommon path from the measurement setup \cite{calosso2014frequency}. To address the temperature fluctuations of the remaining out-of-loop path mainly coming from the two-way optical frequency comparison setup, all the components at each site were housed in a two-layer active temperature controlled copper box with a stability of $0.01$ K. Using these techniques, measurements with the back-to-back system show that noise in our stabilization system contributes namely, $2\times10^{-16}$ at 1 s, reaching $2\times10^{-20}$ after 10,000 s. Compared to the scheme in \cite{hu2020multi}, the achieved long-term stabilities at 10,000 of the noise floor and 145 km fiber link are, respectively, improved by almost factors of 100 and 10.

The article is organized as follows. We illustrate the concept of coherence transfer of laser light through a branching fiber link with passive optical phase stabilization in Sec. \ref{sec2}. We discuss the experimental set-up and experimental results in Sec. \ref{sec3}.  Finally, we conclude in Sec. \ref{sec6} by summarizing the  results. 


\section{Concept of optical frequency dissemination on a branching network}
\label{sec2}

Figure \ref{fig1} shows coherence transfer of our enhanced passive optical frequency transfer on a branching fiber network,  where the optical phase noise induced by environmental perturbations via the fiber link is passively compensated at the remote sites \cite{schediwy2013high, wu2016coherence, ma2018delay}. The laser light, namely optical reference, is coupled into a branching topology fiber network and simultaneously distributes to all remote users. For each remote site, taking the remote site 1 as an example, the light in the fiber link retro-reflected multiple times between the Faraday mirrors (FRMs) independently installed at the local and remote sites. To distinguish the reflected light by the FRM from the stray reflected light along the fiber, the  AOM1 and AOM2 with different driving frequencies are adopted at the local and remote sites, respectively. At the remote site 1,  the remote AOM2 is uniquely used to discriminate its frequency, and also used to correct the phase noise of the single-pass light. Consequently, the different frequencies of the light enable remote users at different sites to compensate for the fiber phase noise independently. The beatnote is obtained onto a PD by heterodyne  beating the single-pass light against the triple-pass light, acting as a phase detector that extracts the phase noise introduced by the fiber link. The detected phase noise is fed into the the remote AOM2 for compensating the phase noise. The detailed working principle of the proposed scheme is as follows. The electric field of the reference light from an optical reference is,
\begin{equation}
E_{0}\propto\cos(\nu t+\phi_s),
\end{equation}
where $\nu$ and $\phi_s$ are the angular frequency and the phase of the reference light. For convenience, we ignore the signal amplitude in the text. The light is passed through a local AOM1 (upshifted mode) working at the angular frequency of $\omega_l$ with the initial phase of $\phi_l$ and then injected into a branching topology fiber network. After passing through the fiber link and the remote AOM2 (downshifted mode)  driven by a radio frequency (RF) signal $E_{r}$ with the angular frequency of $\omega_r$, the output light can have a form of,
\begin{equation}
E_{1}\propto\cos[(\nu+\omega_{l}-\omega_{r})t+\phi_s+\phi_l+\phi_p],
\end{equation}
where $\phi_{p}$ is the phase noise from the fiber introduced by environmental perturbations.  Assuming that the optical phase noise introduced by the environmental perturbations on the fiber link in either direction is equal, the light after transferring through the fiber between the local site and the remote site 1 for another round-trip can be expressed as,
\begin{equation}
E_{2}\propto\cos[(\nu+3(\omega_{l}-\omega_{r}))t+\phi_s+3(\phi_p+\phi_l)].
\end{equation}

By  heterodyne detection between $E_{1}$ and $E_2$ onto the PD1, the lower sideband is filtered out by using a narrow bandwidth RF bandpass filter, obtaining,
\begin{equation}
E_{3}\propto\cos[2(\omega_{l}-\omega_{r})t+2(\phi_p+\phi_l)].
\end{equation}

Then we divide the angular frequency of $E_3$ with a factor of 2, resulting in,
\begin{equation}
E_{4}\propto\cos((\omega_{l}-\omega_{r})t+(\phi_p+\phi_l)).
\end{equation}

Afterwards we combine $E_{4}$ and $E_r$ to drive the RF port of the AOM2. After being compensated, the forward light at the remote site 1 is expressed as, 
\begin{equation}
\begin{split}
E_5&\propto\cos(\nu t+\phi_s)+\sum_{n=1}^{\infty}\cos\left[(\nu+2n(\omega_{l}-\omega_{r}))t\right.\\
&\,\,\,\,\,\,\,\,\,\,\,\,\,\,\,\,\,\,\,\,\,\,\,\,\,\,\,\,\,\,\,\,\,\,\,\,\,\,\left.+\phi_s+2n(\phi_p+\phi_l)\right],
\end{split}
\label{eq7}
\end{equation}
where $n$ represents multiple-trip optical signals between the local site and the remote site 1. The optical frequency as the first term illustrated in  Eq. \ref{eq7} is independent of the phase noise resulting from phase perturbations on the fiber link and the phase noise coming from the remote RF reference. Compared to the scheme presented in \cite{hu2020multi}, here only one  AOM at the remote site is simultaneously used as a frequency distinguisher for distinguishing its unique frequency from other sites' and avoiding the stray light from the desirable signal, and as an optical actuator for compensating the phase noise coming from the optical fiber. With this configuration, it significantly reduces the out-of-loop uncommon path, resulting in the improvement of the system performance in particular the long-term stability as discussed in the Sec. I.



By taking into account the fiber propagation delay, the residual phase noise power spectral density (PSD) at any remote site in terms of the single-pass free-running phase noise PSD, $S_{\text{fiber}}(\omega)$, and the propagation delay of the fiber, $\tau_0$, can be calculated as \cite{hu2020multi,williams2008high},
\begin{equation}
S_{\text{remote}}(\omega)\simeq\frac{7}{3}(\omega\tau_0)^2S_{\text{fiber}}(\omega).
\label{eq18}
\end{equation}




\section{Experimental setup and results}
\label{sec3}




\subsection{Experimental setup}
\begin{figure}[htbp]
\centering
\includegraphics[width=1\linewidth]{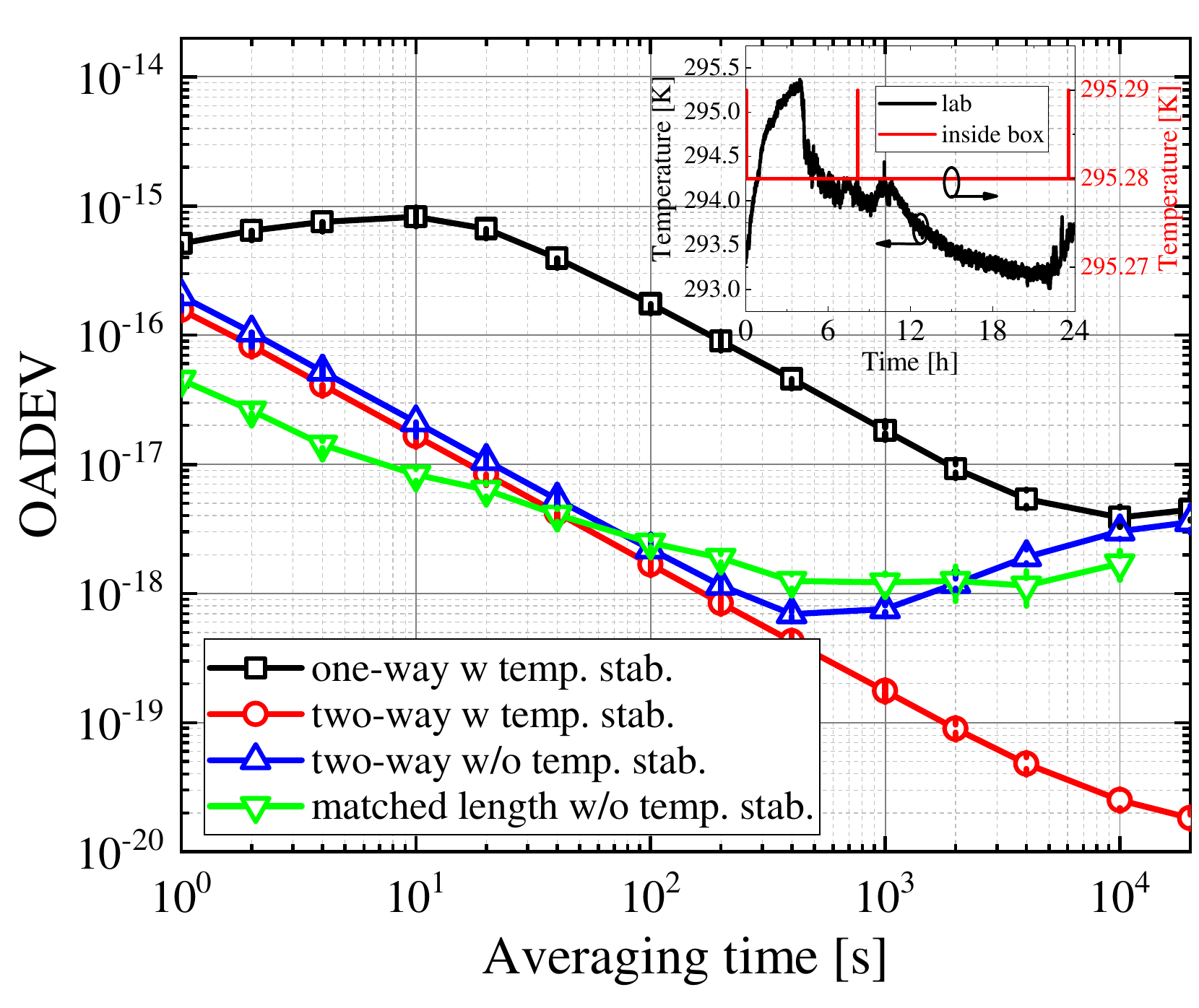}
\caption{Measured fractional frequency instabilities of the interferometer by short connecting the local site and the remote site 1 with the different end-to-end stability measurement methods for  the one-way optical frequency comparison acquired the beatnote at the local site (black squares) with the temperature stabilized interferometer,  the two-way optical frequency comparison with the temperature stabilized interferometer (red circles), the two-way optical frequency comparison without the temperature stabilized interferometer (blue up triangles) and the configuration in which the two sites are connected by an equivalent arm length $1\times2$ optical coupler without the temperature stabilized interferometer (green down triangles). The inset shows that with the implementation of the temperature stabilization, the peak-to-peak temperature fluctuations can be effectively reduced from 2.3 K to 0.01 K.}
\label{fig2}
\end{figure}

We have demonstrated the proposed  technique with the experimental configuration as shown in Fig. \ref{fig1} with the two remote sites (remote site 1 and remote site 2). The optical reference is a narrow-linewidth optical source (NKT X15) at a frequency near 193 THz with a linewidth of 100 Hz. The signal was transmitted along a 145 km fiber link to the remote site 1 and 50 km to the remote site 2. The AOM1, AOM2 and AOM3 are working at the upshifted, downshifted and downshifted modes, respectively. Here  we set $\omega_l=2\pi\times75$ MHz,  $\omega_{r}=2\pi\times45$ MHz (remote site 1) and $\omega_{r,2}=2\pi\times93$ MHz (remote site 2). Due to the limitation of the bandwidth of the available AOMs, after detecting the phase noise of each fiber link, we mix the beatnote with an assistant frequency of 5 MHz (101 MHz) and extract the upper (lower) sideband with a frequency of 35 MHz (83 MHz) for the remote site 1 (2). All these RF frequencies are provided by a direct-digital-synthesizer (DDS) generator, phase locked to a 10 MHz rubidium clock. With this configuration, we can have  an out-of-loop beatnote of $40$ MHz and 8 MHz for the remote site 1 and 2, respectively.  Here to ensure its long-term transfer performance due to the strong laboratory temperature variations with a peak-to-peak value of 2 $^{\circ}$C (shown in Fig. \ref{fig2}), we install one manual fiber polarization controller (FPC) at the end of each branching fiber link. We typically have  to manually adjust the polarization approximately per day. In the next step, an automatic polarization tracker is planned to be installed to reduce the polarization effect  \cite{wang2020novel}.

To reduce the instability floor caused by the laboratory temperature variations, all the optical components at each site as the red shaded areas shown in Fig. \ref{fig1} including FRMs, isolators and a few couplers are housed in a two-layer copper box. The external dimensions of the layers are {$200\times200\times100$ mm and $170\times170\times70$ mm}. Each copper layer has the thickness of 10 mm covered by a thick insulation wool layer with the thickness of 5 mm for acoustic noise and passive temperature shielding.  The temperature of each copper layer is actively stabilized around 295 K using Peltier elements, and a digital temperature control circuit based on a micro-controller  is used to stabilize the temperature fluctuations. 

In order to effectively measure the transfer stability at each remote site, the outputs of the local and each remote sites are connected with a short fiber link, allowing for the measurement of the end-to-end frequency fluctuations of the link using a two-way frequency comparison method as illustrated the dashed boxes in Fig. \ref{fig1}. The beatnote signal between the local (remote) site output and the remote (local) site output is detected on the PD2 (PD3) at the local (remote) site. We can acquire the frequency difference between the output signals of the two sites without any effect coming from the short fiber connecting the outputs of the local and remote sites. We use non-averaging $\Pi$-type frequency counters, which are referenced to the RF frequency sourcing from the DDS to record the beating frequency between the fiber input light and the output light. At the same time, the phase noise is measured with a fast Fourier transform (FFT) analyzer \cite{hu2020cancelling}.

\subsection{Interferometer noise floor characterization}

We characterize the interferometer noise floor by short connecting two sites and measure the end-to-end instability with different configurations. Figure \ref{fig2} illustrates the corresponding fractional frequency instability. Measurements of the frequency instability are calculated by the OADEV and taken with the counters operating in $\Pi$-averaging type with 1 s gate time \cite{dawkins2007considerations}. The inset of Fig. \ref{fig2} shows  the typical temperature stabilization performance, demonstrating that the peak-to-peak temperature fluctuations can be effectively suppressed from 2.3 K  to 0.01 K with a factor of 230. Without the temperature stabilized optical paths of the outside loop for the end-to-end stability measurement performed by  the one-way optical frequency comparison acquired the beatnote at the local site, the two-way optical frequency comparison and the configuration in which the two sites are connected by an equivalent arm length $1\times2$ optical coupler without the temperature stabilized interferometer, the long-time stability in terms of the OADEV is constrained at a few times $10^{-18}$, which is similar with our previous results \cite{hu2020multi, tian2020hybrid}. The limit can be effectively suppressed by two orders of magnitude by adopting the two-way optical frequency comparison method with the temperature stabilized interferometer. We regard the observed  instability around of $2\times 10^{-16}$ at 1 s and $2\times 10^{-20}$ at 10,000 s in our present setup. We can conclude that the temperature stabilization of the interferometer could  improve the long-term stability and the two-way optical frequency comparison method can be effectively used to evaluate the end-to-end performance.

\subsection{Frequency spectrum at the remote site}

\begin{figure}[htbp]
\centering
\includegraphics[width=1\linewidth]{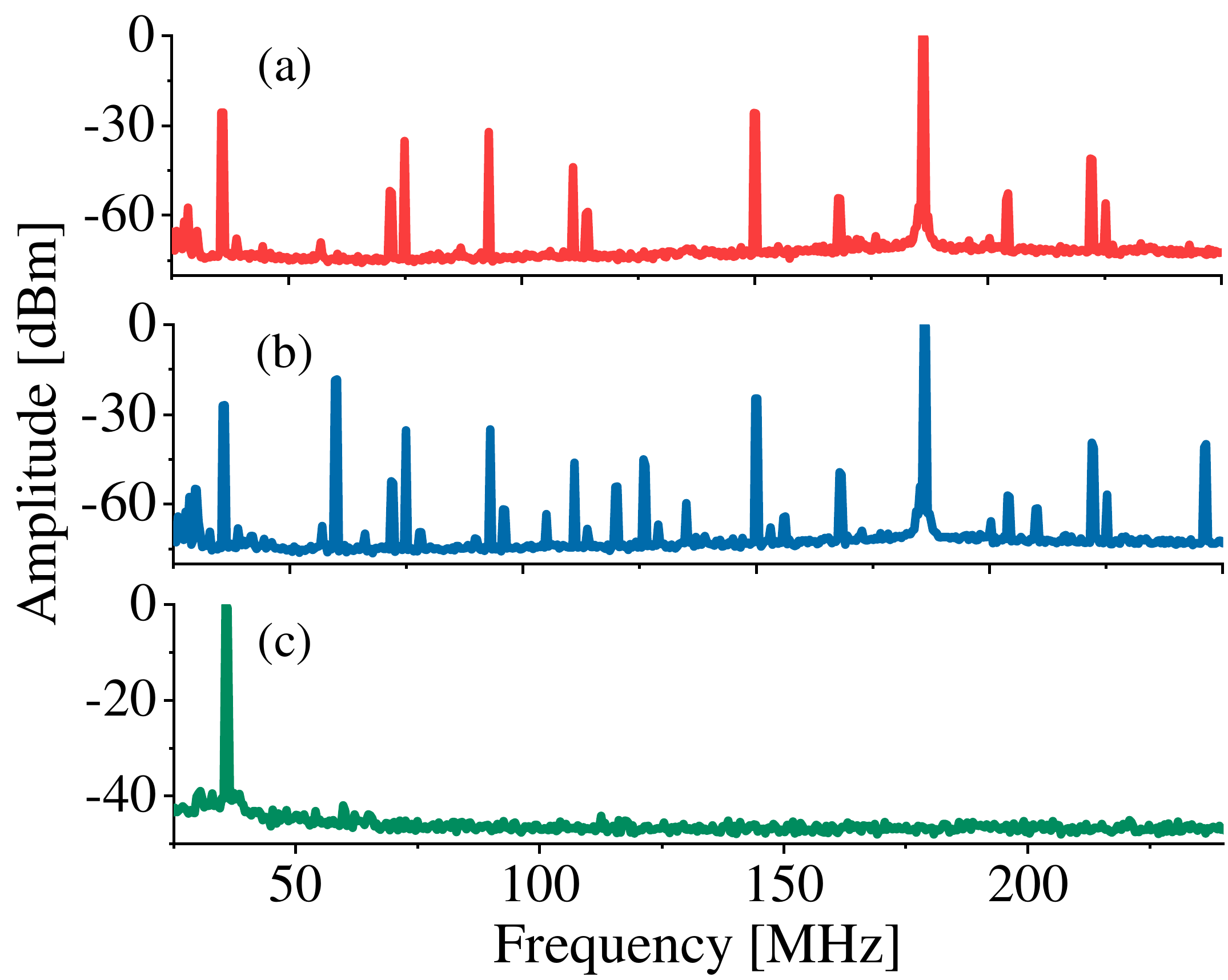}
\caption{Normalized power spectrum as recovered by the PD at the remote site 2 for the remote 1 disconnected (a), for the remote 1 connected (b),  and for the remote 1 connected, but after a bandpass filter centered on 36 MHz and an amplifier (c), which suppresses all unwanted signals by at least 40 dB below the desirable signal.}
\label{fig3}
\end{figure}


Figure \ref{fig2} (a), (b) and (c), respectively, show the normalized power spectrum detected by the PD at the remote site 2 for the remote 1 disconnected, for the remote 1 connected,  and for the remote 1 connected, but after a bandpass filter centered on 36 MHz and an amplifier, which suppresses all unwanted signals by at least 40 dB below the desirable signal. The unwanted signals could be further rejected in the mixing process and by a low-pass filter at the output of the mixer \cite{schediwy2013high}. Simultaneously, the remote site 1 with the frequency of 60 MHz could also be effectively filtered out by adopting an RF bandpass filter.

\subsection{Frequency-domain characterization at the remote sites}

\begin{figure}[htbp]
\centering
\includegraphics[width=1\linewidth]{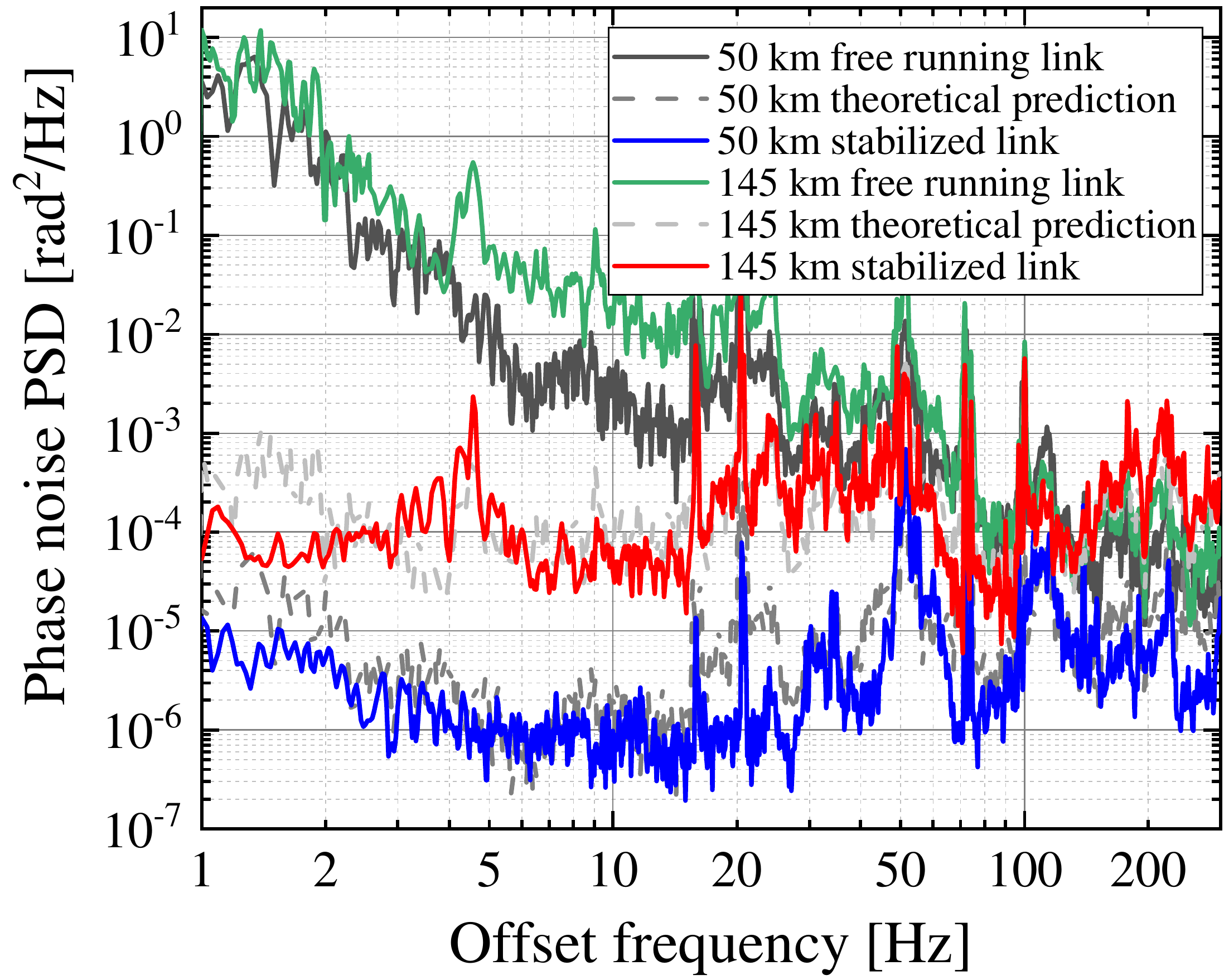}
\caption{Measured phase noise PSDs of the  free-running fiber link for the remote site 1 (green curve) and the remote site 2 (black curve) and the stabilized link with passive phase noise cancellation  for the remote site 1 (red curve) and the remote site 2 (blue curve). Strong servo bumps can be effectively eliminated in the passive phase noise cancellation scheme. The dashed curves are the theoretical prediction based on Eq. \ref{eq18}.}
\label{fig4}
\end{figure}

To characterize the frequency domain performance, we measured the phase noise PSDs of  the remote site 1 over 145 km and the remote site 2 over 50 km  for both the stabilized and the unstabilized configurations as plotted in Fig. \ref{fig4}. Both remote sites for the free running fiber links are very similar and typical for optical fiber links, with the noise dependence of  $h_0f^{-2}$, illustrating that the phase noise of the free-running fiber is mainly limited by the flicker phase noise. The compensated phase noise PSDs for the remote site 1 and 2 are, respectively,  close to  $10^{-4}$ rad$^2$/Hz  and $10^{-5}$ rad$^2$/Hz in the low frequency range of several hertzs, illustrating that the compensated fiber link is mainly constrained by the white phase noise after passive phase noise compensation. Noise correction bandwidths are limited by the fiber propagation delay with the bandwidth of 195 Hz ($1/(4\sqrt{7}\tau_0)$) and 567 Hz for the remote site 1 and 2, respectively. We checked that the noise floors of both outputs were below these PSDs. The noise rejections for both remote sites are compatible with the theoretical limit calculated with Eq. \ref{eq18} as the dashed curves shown in Fig. \ref{fig4}, demonstrating that the phase noise rejection is optimized.  Same with all the passive phase noise cancellation schemes \cite{hu2020multi, hu2020cancelling, hu2020passive}, strong servo bumps have been effectively eliminated.


\begin{figure}[htbp]
\centering
\includegraphics[width=1\linewidth]{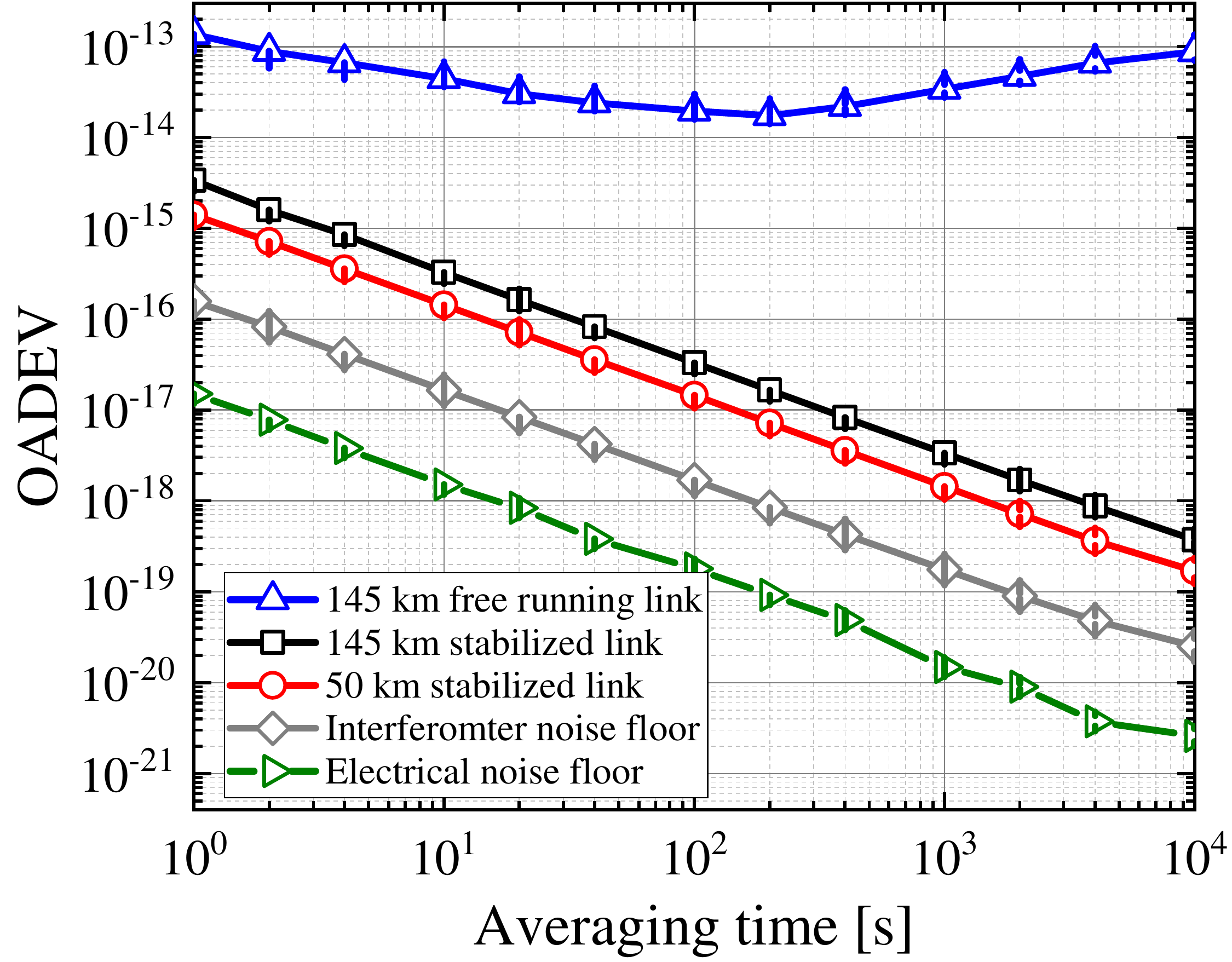}
\caption{Measured fractional frequency instabilities in terms of the OADEV of the free-running fiber link for the remote site 1 (blue triangle) and the stabilized link for the remote site 1 (black square) and the remote site 2 (red circle). The measured  interferometer noise floor and  electrical noise floor are also shown by gray diamonds and green right triangles, respectively.}
\label{fig5}
\end{figure}

\subsection{Time-domain characterization at the remote sites}

Figure  \ref{fig5}  illustrates the time-domain characterization of  the frequency stability in terms of the OADEV.  With the implementation of  fiber noise cancellation for the remote site 1 (2), optical frequency transfer achieves a fractional frequency stability of $3.4\times10^{-15}$ ($1.4\times10^{-15}$) at the integration time of $1$ s, decreases and reaches $3.7\times10^{-19}$ ($1.7\times10^{-19}$) at $10, 000$ s. In our experiment, we observe that the stability of optical frequency transfer is improved by at least four orders of magnitude at the integration time of 10,000 s by activating the phase noise cancellation setup. With the optimized configuration by reducing the out-of-loop optical path based on the configuration in \cite{hu2020multi}, the stability up to 10,000 s is rather than limited by the system noise floor \cite{hu2020multi, hu2020passive} and the noise floor at 10,000 s is improved by almost one order of magnitude. Once the fiber noise cancellation setups are engaged, the frequency fluctuations can be effectively suppressed and no longer dominate the instability of the optical signals at both remote sites. As illustrated in Fig. \ref{fig4}, the performance reaches the theoretical limit and is at the same level as the state of the art results \cite{bercy2016ultrastable}. As a comparison, we also measured the electrical noise floor in which we only keep the electrical components as the right triangles shown in Fig. \ref{fig5}, demonstrating the electrical noise floor is one order magnitude lower than the interferometer noise floor.




\subsection{Frequency transfer accuracy}

\begin{figure}[htbp]
\centering
\includegraphics[width=1\linewidth]{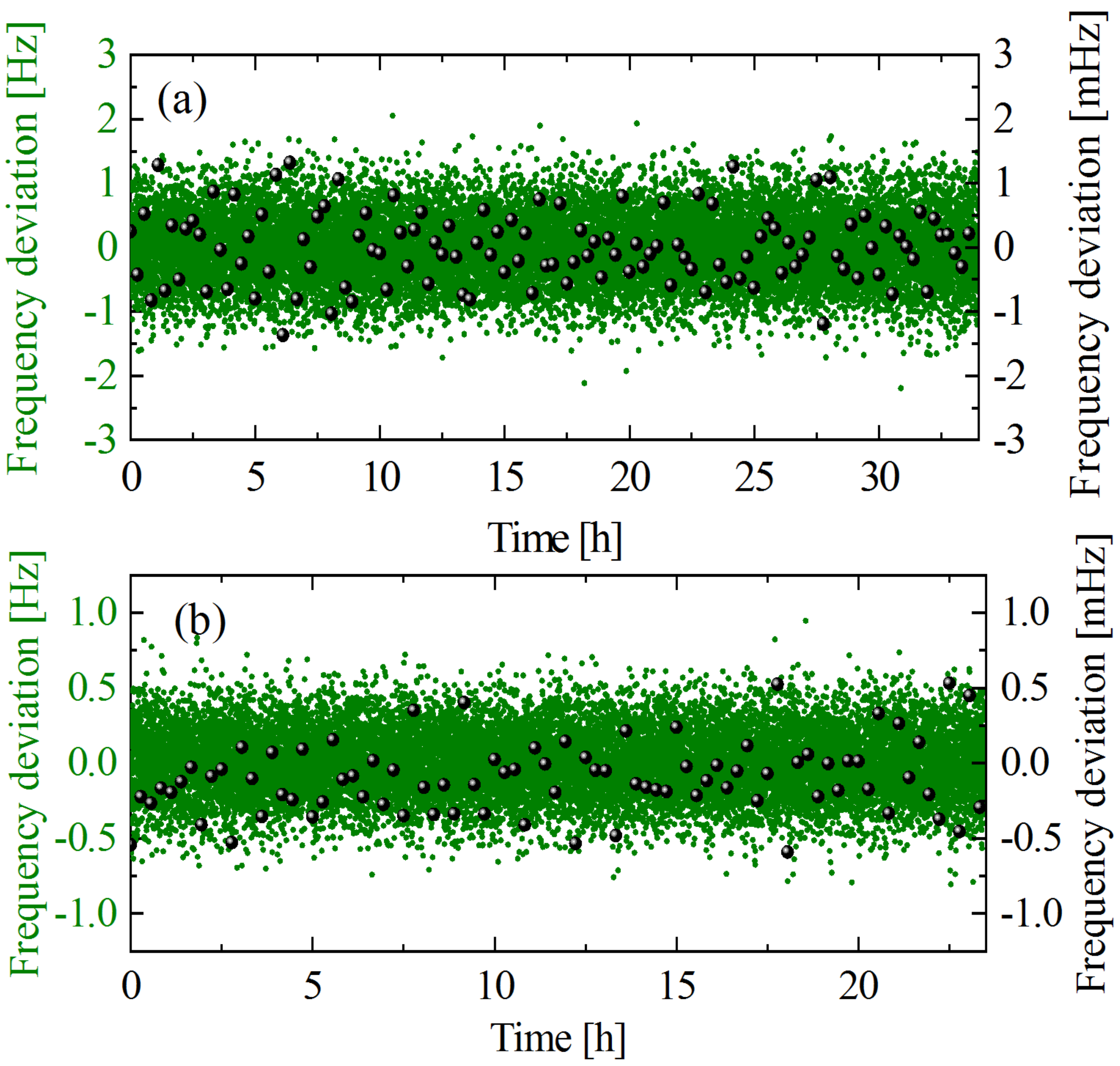}
\caption{ Frequency deviation between local and transferred frequencies of the remote site 1 (a) and the remote site 2 (b). Data were acquired with a dead-time free $\Pi$-type frequency counter with 1 s gate time (green points, left axis). We calculated unweighted mean ($\Pi$-type) values for all cycle-slip free 1,000 s long segments (black dots, right frequency axis, enlarged scale). The total 122,927 and 85,253 $\Pi$-type counter data were taken for the remote site 1 (a) and the remote site 2 (b), respectively.}
\label{fig6}
\end{figure}

Complementary to the characterization of the system performance in the frequency and time domains, we also performed an evaluation of the accuracy of optical frequency transfer at both remote sites, which cannot be identified in the instability evaluation in Fig. \ref{fig5}. Figure \ref{fig6}(a) and (b), respectively, show the frequency deviations of the measured data for the the remote site 1 and 2, recorded with 1 s gate time and $\Pi$-type counters. For the remote site 1 as illustrated in Fig. \ref{fig6}(a), the data were measured over successive 122,927 s (green point, left axis). By averaging the data every 1000 seconds, we can have 122 points as shown in the right axis of Fig. \ref{fig6}(a).  The 122 points have the arithmetic mean and standard deviation of 8.32 $\mu$Hz ($4.3\times10^{-20}$) and $0.55$ mHz ($2.9\times10^{-19}$), respectively. The standar deviation is a factor of 1,000 smaller than the OADEV at 1 s. By taking into account the long-term stability of optical frequency transfer as illustrated  in Fig. \ref{fig5}, we conservatively constrain the accuracy of optical frequency transfer  as shown in the last data point of the {OADEV}, resulting in a relative frequency accuracy of $3.7\times10^{-19}$. By adopting the same procedure, the mean frequency offset for the remote site 2 was calculated using the total 85,253 $\Pi$-type counter data to be -985 $\mu$Hz ($-5.1\times10^{-19}$). We divide all the data into 85 groups, and calculate a mean value for each group. Based on those mean values, we calculate the standard deviation of the 85 points is $0.24$ mHz ($1.2\times10^{-18}$). As the long-term {OADEV} at 10,000 s of the data set for the remote site 2  is constrained at $1.7\times10^{-19}$, we estimate that the mean frequency offset is $-5.1\times10^{-19}$ with a statistical uncertainty of $1.7\times10^{-19}$ for the remote site 2. We can conclude that there is no systematic frequency shift arising in the extraction setup at a level of a few $10^{-19}$.




\section{conclusion}
\label{sec6}


In conclusion, we demonstrated a novel  technique for simultaneous dissemination of high-precision optical-frequency signals to multiple independent locations with passive phase noise cancellation. Each remote AOM is simultaneously taken as a frequency distinguisher for distinguishing its unique frequency from other sites' and as an optical actuator for compensating the phase noise coming from the optical fiber. By implementing this configuration, we successfully incorporate the out-of-loop component, namely a fiber-pigtailed AOM, into the loop, enabling that the AOM became part of the stabilized fiber link and the phase noise coming from it can be effectively suppressed. In comparison with multiple-user optical frequency transfer over a ring and a bus fiber links,  in which the phase noise stabilization of each user end is strongly dependent on the main fiber link, high-performance multiple-user optical frequency transfer demonstrated here could increase the adaptability of optical frequency transfer to the existing fiber network. At the same time, it could also improve the reliability of the network because the phase noise compensation of each user node is independent of each other. By implementing the active temperature stabilization system and the two-way optical frequency comparison for evaluating the system performance, the back-to-back system shows that the noise floor in our stabilization system contributes fluctuations that are $2\times10^{-16}$ at 1 s, reaching $2\times10^{-20}$ after 10,000 s. With implementing these techniques, we demonstrated transfer of a laser light through a branching fiber network with 50 km and 145 km two fiber links  simultaneously and found no systematic offset between the local  and remote frequencies within the uncertainty of a few $10^{-19}$. This technique with passive phase noise compensation could significantly shorten the response speed and phase recovery time of optical frequency transfer, enabling a wide range of applications beyond metrology over reliable and scalable branching fiber networks.



\ifCLASSOPTIONcaptionsoff
  \newpage
\fi



%


\begin{IEEEbiographynophoto}{Ruimin Xue} received the B.S. degree from Hunan University, China, in 2018. She is currently a graduate student in the State Key Laboratory of Advanced Optical Communication Systems and Networks, Department of Electronic Engineering, Shanghai Jiao Tong University. Her current research interests include photonic signal transmission.
\end{IEEEbiographynophoto}

\begin{IEEEbiographynophoto}{Liang Hu}
received the B.S. degree from Hangzhou Dianzi University, China, in 2011, and the M.S. degree from Shanghai Jiao Tong University, China, in 2014. He received the Ph.D. degree from University of Florence, Italy, in 2017 during which he was a Marie-Curie Early Stage Researcher at FACT project. He is currently a Tenure-Track Assistant Professor in the State Key Laboratory of Advanced Optical Communication Systems and Networks, Department of Electronic Engineering, Shanghai Jiao Tong University, China. His current research interests include photonic signal transmission and atom interferometry.
\end{IEEEbiographynophoto}


\begin{IEEEbiographynophoto}{Jianguo Shen}
received his Bachelor Degree on Physics Educations from Zhengjiang Normal University in 2002, Master Degree on Circuit and system from Hangzhou Dianzi University in 2007, Ph.D on Electromagnetic and Microwave Technology from Shanghai Jiaotong University in 2015. Currently, he is an associate professor at Zhejiang normal university of china. His research interests include microwave photonic signals processing and time and frequency transfer over the optical fiber.   
\end{IEEEbiographynophoto}

\begin{IEEEbiographynophoto}{Jianping Chen}
received the B.S. degree from Zhejiang University, China, in 1983, and the M.S. and Ph.D. degrees from Shanghai Jiao Tong University, China, in 1986 and 1992, respectively. He is currently a Professor in the State Key Laboratory of Advanced Optical Communication Systems and Networks, Department of Electronic Engineering, Shanghai Jiao Tong University. His main research interests include opto-electronic devices and integration, photonic signal processing, and system applications. He is a Principal Scientist of National Basic Research Program of China (also known as 973 Program).
\end{IEEEbiographynophoto}

\begin{IEEEbiographynophoto}{Guiling Wu}
received the B.S. degree from Haer Bing Institute of Technology, China, in 1995, and the M.S. and Ph.D. degrees from Huazhong University of Science and Technology, China, in 1998 and 2001, respectively. He is currently a Professor in the State Key Laboratory of Advanced Optical Communication Systems and Networks, Department of Electronic Engineering, Shanghai Jiao Tong University, China. His current research interests include photonic signal processing and transmission.
\end{IEEEbiographynophoto}

\end{document}